# Steering Selective Formation and 2D Crystallization of [4]Radialenes on Au(111) via [1+1+1+1] Cycloaddition of Isocyanides and Enantioselective Molecular Recognition

Jian-Wei Liu, Ying Wang*, Cui-Ping Wu, Jia-Xin Li, Li-Xia Kang, Jian-Hui Fu, Wen-Wen Gong, Pei-Nian Liu*, and Deng-Yuan Li*

*State Key Laboratory of Natural Medicines, School of Pharmacy, China Pharmaceutical University, Nanjing, 211198, China*

*School of Chemistry and Molecular Engineering, East China University of Science & Technology, Shanghai 200237, China*

*ChangXin Memory Technologies, Inc. (CXMT)No. 799, Qide Road, Economic and Technological Development Area, Hefei, 230601, China*

**TOC graphic**

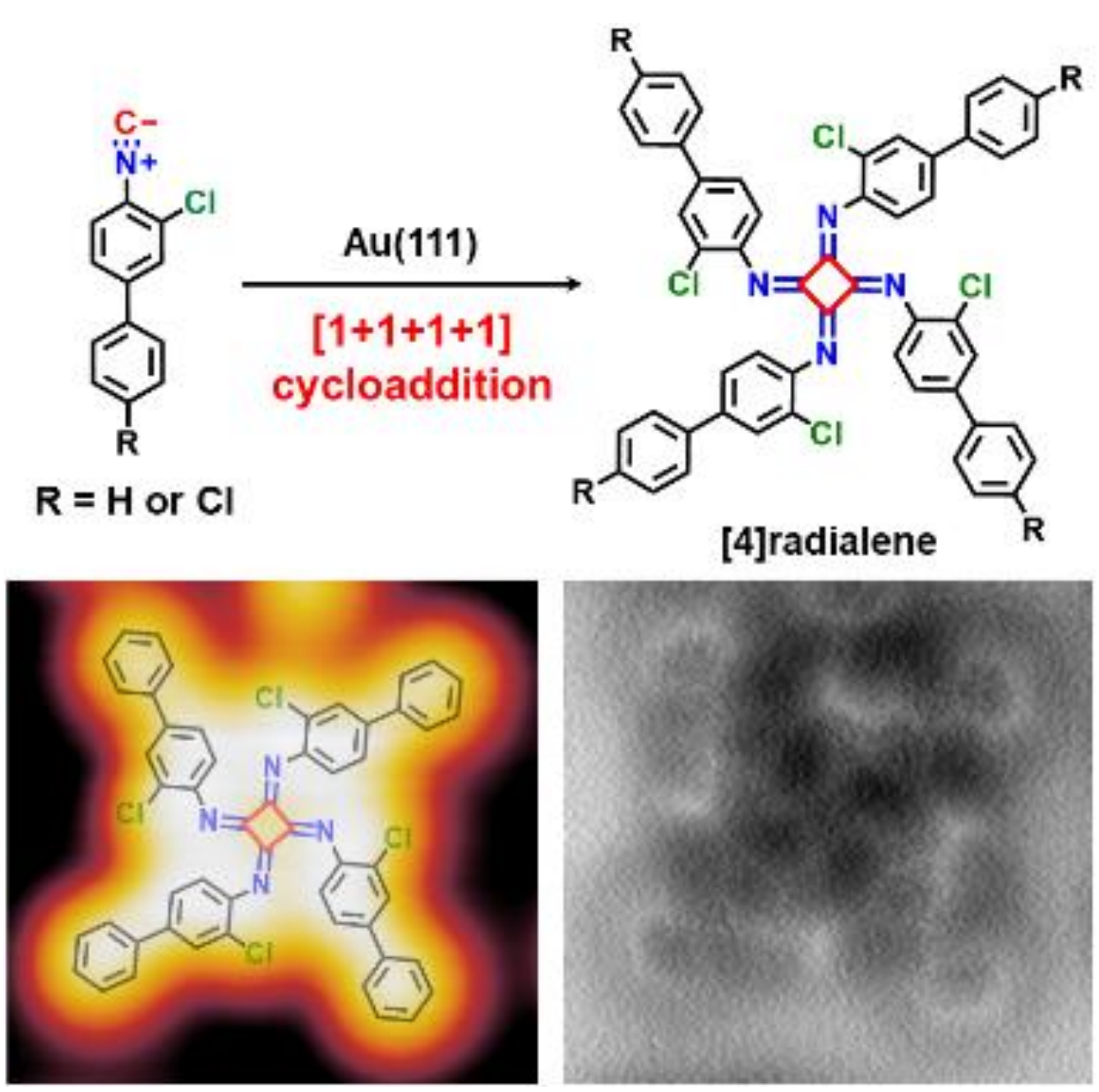

## Abstract

Conjugated carbon rings are fundamental skeletons of organic functional materials, and their selective formation is of paramount importance in molecular materials engineering. However, steering the formation and 2D crystallization of conjugated carbon rings on the surface with high chemo- and stereoselectivities remains a great challenge. Here, we report a highly chemoselective [1+1+1+1] cycloaddition of isocyanides on the Au(111) surface, which affords the stereospecific tetraaza[4]radialene products and further enables their long-range-ordered 2D crystallization via enantioselective molecular recognition. Using the progressive annealing method, we found that at room temperature, isocyanides undergo a coordination reaction with Au adatoms to form two-fold symmetric isocyanide-Au-isocyanide complexes. In contrast, gradually increasing the annealing temperature induces the transformation of these complexes and subsequent covalent polymerization, leading to the selective generation of tetraaza[4]radialenes with homotactic configurations. The tetraaza[4]radialenes further assemble into 2D homochiral molecular crystals through enantioselective molecular recognition driven by multiple C–H···Cl hydrogen-bonding interactions. By combining scanning tunneling microscopy/spectroscopy and non-contact atomic force microscopy, we determined the atomic structure and molecular orbitals of tetraaza[4]radialene, confirming that its four-membered ring adopts a planar geometry with a localized lowest unoccupied molecular orbital. Density functional theory calculations suggest that the [1+1+1+1] cycloaddition process involves stepwise formation of C–C bonds and its high selectivity arises from the spatial steric hindrance. Our findings provide new insights into the selective formation of conjugated rings on surfaces and have implications for engineering 2D homochiral molecular crystallization.

## Introduction

Conjugated carbon rings are fundamental skeletons of organic functional materials. Their selective formation is of paramount importance in modern synthetic chemistry and molecular materials engineering, as the intrinsic structure parameters—including ring size and topology—directly govern the electronic delocalization and aromatic conjugation, thereby critically determining the physicochemical properties of the resulting molecular materials. Over the past two decades, on-surface synthesis has provided a unique pathway for the atomically precise fabrication of conjugated carbon rings and their extended structures,[1–4] including graphene nanoribbons,[5–12] triangulenes,[13–15] cyclocarbons,[16–18] and biphenylene sheets.[19] Moreover, a variety of conjugated carbon rings at the single-molecule scale have been explored and visualized through scanning probe microscopy,[20–37] providing valuable experimental insight for understanding molecular aromatic/anti-aromatic[20–24] and electrical/magnetic properties.[25–37] Generally, the formation of conjugated carbon rings on the surface mainly includes direct cyclization involving the cycloaddition of unsaturated carbon–carbon[38–43] or carbon–nitrogen bonds,[44–47] and indirect cyclization involving debromination,[48–52] dehydrogenation,[53–55] and dehydration cycloaddition reactions.[56,57] Nevertheless, precisely steering the formation and two-dimensional (2D) crystallization of conjugated carbon rings with high chemo- and stereoselectivities on the surface remains a formidable challenge.

[n]Radialenes (n = the number of exocyclic double bonds) are unique carbocyclic structures possessing three or more cross-conjugated exocyclic double bonds.[58] [3]Radialene is a cross-conjugated six-center, six π-electron system, and [4]radialene is a cross-conjugated eight-center, eight π-electron system, endowing them with different physicochemical properties. Recently, we achieved the first synthesis of triaza[3]radialene and its 2D crystals on Ag(111) via selective [1+1+1] cycloaddition of isocyanides (Figure 1a, Path I).[45,59] Li et al. reported the first synthesis of tetraphenyl[4]radialene on Cu(100) via one-step cycloaddition of phenylacetylenes.[60] However, the heteroatom-doped [4]radialenes and their 2D crystals on the surface have not been reported until now, owing to the lack of a suitable synthetic method.

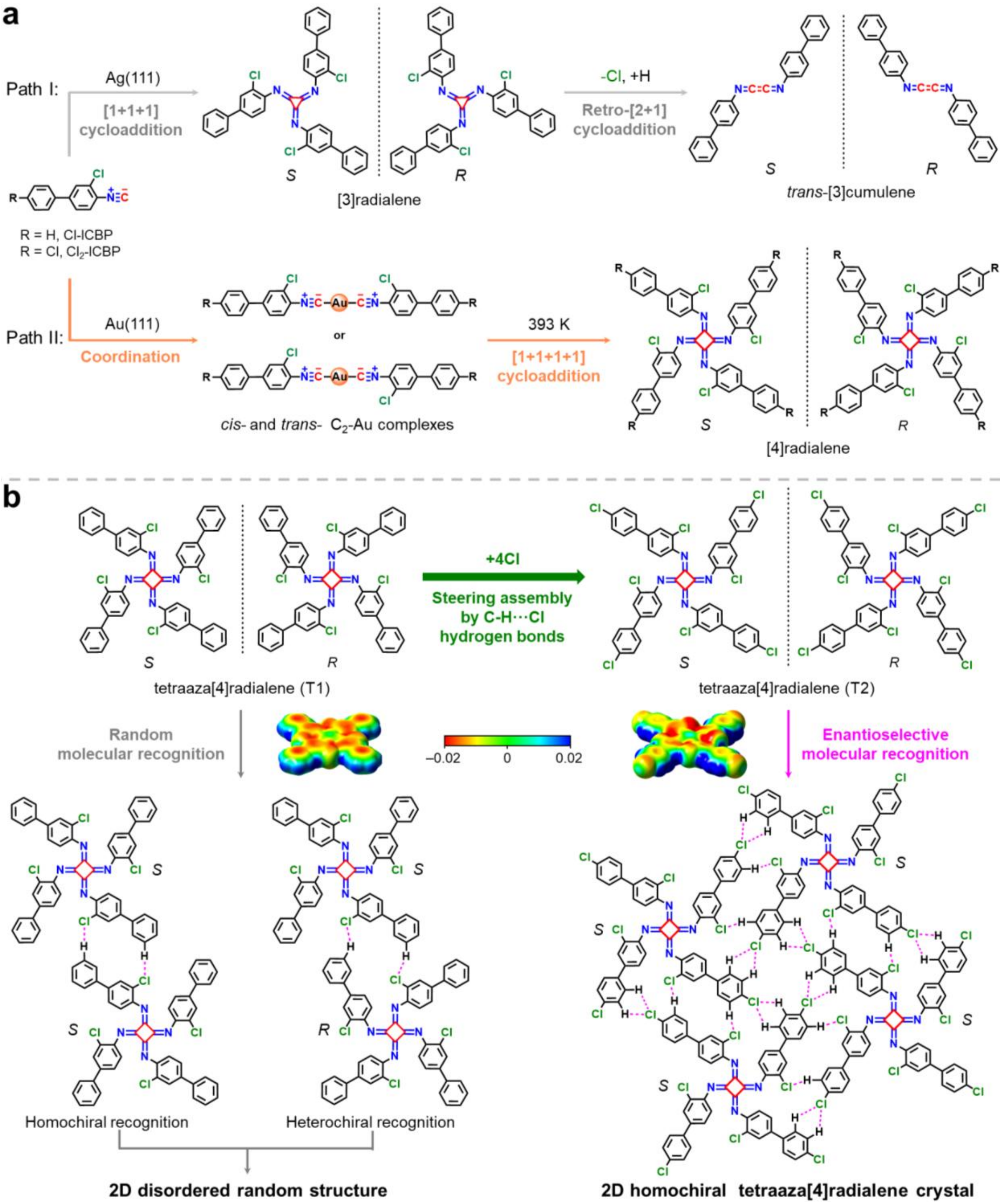

**Figure 1.** Steering selective formation and 2D crystallization of radialenes from isocyanides on metal surfaces. (a) Path I: Transformation from isocyanides to [3]radialenes to *trans*-[3]cumulenes on the Ag(111) surface via cascade reaction of [1+1+1] and retro-[2+1] cycloadditions. Path II: Transformation from isocyanides to [4]radialenes on the Au(111) surface via [1+1+1+1] cycloaddition. (b) Steering the enantioselective molecular recognition of tetraaza[4]radialenes via multiple Cl atoms-derived C–H···Cl hydrogen-bonding interactions. The insets show the DFT-calculated electrostatic potential (ESP) map of *R*-tetraaza[4]radialene (*R*-T1) and *S*-tetraaza[4]radialene (*S*-T2) in the gas phase. The highest and lowest electron densities appear as blue and red regions, respectively.

In this study, we developed a highly chemoselective [1+1+1+1] cycloaddition of isocyanides on the Au(111) surface to give stereospecific tetraaza[4]radialene products. We further achieved their 2D crystallization via enantioselective molecular recognition. At room temperature, isocyanides on Au(111) underwent a coordination reaction with Au adatoms to form two-fold symmetric isocyanide-Au-isocyanide complexes. In contrast, stepwise higher annealing temperatures induced the transformation of these complexes and subsequent covalent polymerization, resulting in the selective generation of tetraaza[4]radialenes with homotactic configurations (Figure 1a, Path II). The tetraaza[4]radialenes further assembled into 2D homochiral molecular crystals through enantioselective molecular recognition based on multiple C–H···Cl hydrogen-bonding interactions (Figure 1b). Combining scanning tunneling microscopy/spectroscopy (STM/STS), non-contact atomic force microscopy (nc-AFM), and density functional theory (DFT) calculations, we identified the atomic structure and molecular orbitals of tetraaza[4]radialenes and demonstrated that the [1+1+1+1] cycloaddition process involves stepwise formation of C–C bonds and that its high selectivity is attributed to the spatial steric hindrance under surface confinement conditions.

## Results and Discussion

### Design for the transformation of isocyanides to [4]radialenes

In our previous work, the precursor 3-chloro-4-isocyano-1,1'-biphenyl (Cl-ICBP) on the Ag(111) surface can undergo [1+1+1] and retro-[2+1] cycloaddition accompanied by dechlorination as the annealing temperature increased, resulting in the transformation from [3]radialenes to [3]cumulenes (Figure 1a, Path I).[59] To control the cyclization of Cl-ICBP on the surface, the slightly inert Au(111) surface was selected as a substrate, which is expected to avoid dechlorination before cyclization. After depositing the precursor Cl-ICBP on a pristine Au(111) surface held at approximately 293 K under ultra-high vacuum (UHV) conditions, STM measurements reveal the formation of large-scale close-packed molecular structures with some regular phases (Figures 2a and S2a). High-resolution STM image (Figure 2d) shows that these close-packed molecular structures are composed of rod-like species containing two gourd-shaped Cl-ICBP molecules and a central Au adatom.[61] Further DFT calculations show that the experimental STM image is consistent with the simulated STM image (Figure S2b-d). These results demonstrate that the precursors Cl-ICBP on Au(111) successfully underwent a coordination reaction with Au adatoms to form two-fold symmetric isocyanide-Au-isocyanide complexes with *cis*- and *trans*-configurations; then these complexes underwent random molecular recognition through C–H···Cl hydrogen-bonding interactions, resulting in the disordered assembly structures with partly ordered phases.

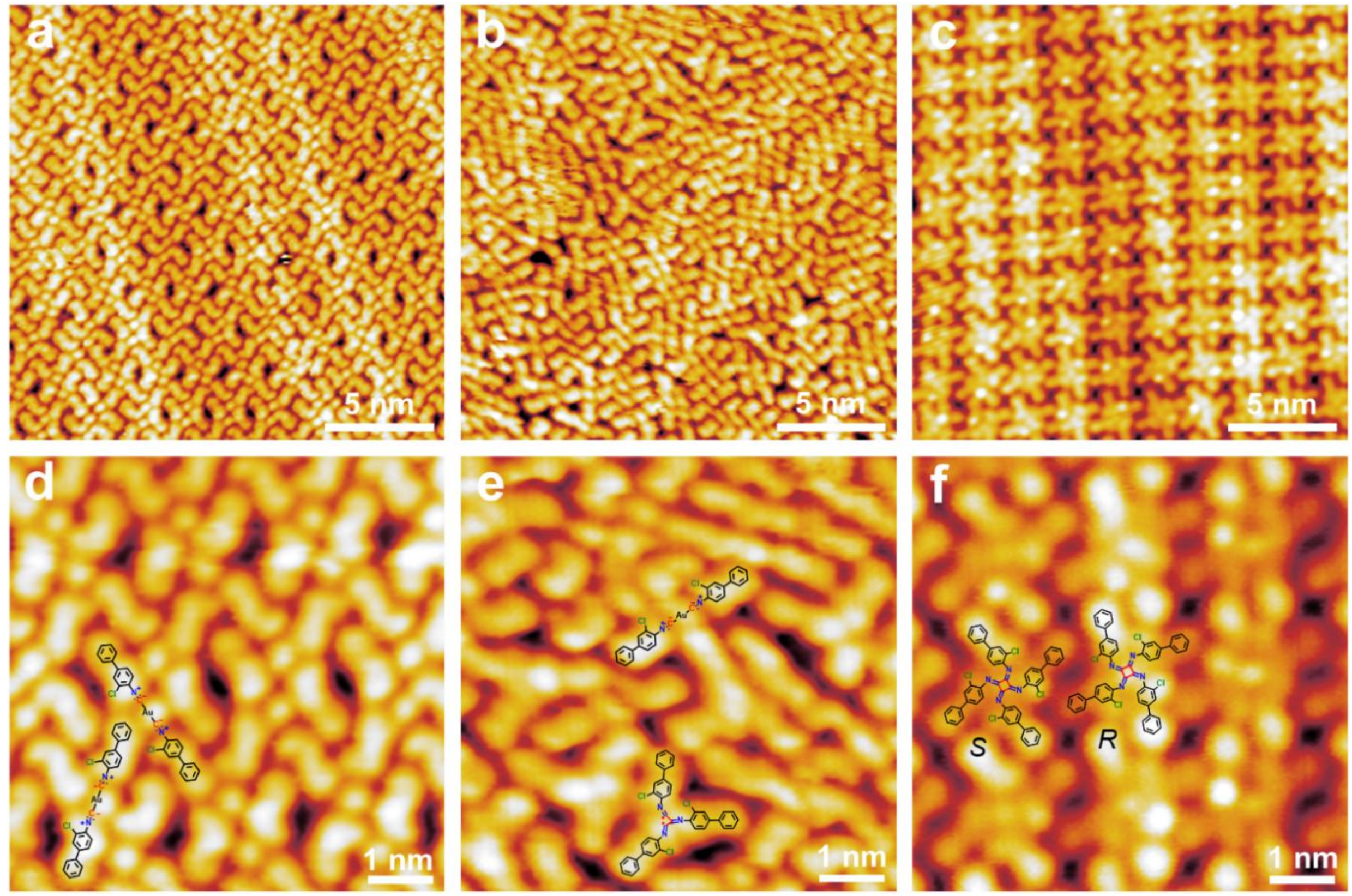

**Figure 2.** Transformation of isocyanides to tetraaza[4]radialenes on Au(111) via selective [1+1+1+1] cycloaddition. (a) High-resolution and (d) zoom-in STM images of coordination assembly of Cl-ICBP. (b) High-resolution and (e) zoom-in STM images after annealing at 373 K. (c) High-resolution and (f) zoom-in STM images of mixed assembly of tetraaza[4]radialenes with two homotactic configurations (*S* and *R*). Scanning parameters: (a) $U$ = -1.63 V, $I$ = 0.55 nA. (b) $U$ = -1.03 V, $I$ = 0.11 nA. (c and f) $U$ = 0.49 V, $I$ = 0.03 nA. (d) $U$ = -1.54 V, $I$ = 0.53 nA. (e) $U$ = -1.03 V, $I$ = 0.12 nA.

Subsequently, the sample containing the coordination structures was annealed at 373 K, and the STM image (Figure 2b) shows that most of the regular coordination structures disappear, generating a disordered phase, in which some Y-shaped species with three leaves are frequently observed. Zoom-in STM image shows that the Y-shaped species are composed of three Cl-ICBP molecules, which are different to the triaza[3]radialenes and similar to their ring-opening intermediates.[59] Upon further annealing to 393 K, a new regular domains appear (Figure 2c), composed of a mixture of two four-leafed species featuring homotactic configurations (*S* and *R*) (Figure 2f). This result suggests that the four-leafed species are *S*- or *R*-tetraaza[4]radialene (*S*-T1 or *R*-T1), formed via a [1+1+1+1] cycloaddition involving four Cl-ICBP molecules. Further repeated annealing at the same temperature leads to an increase in the size of regular domains (Figure S3a). Notably, one of the four leaves in the T1 species appears as a brighter dot; this feature is attributed to the spatial steric hindrance caused by Au(111) surface-confined adsorption, which is supported by DFT simulation (Figure S3b-d).

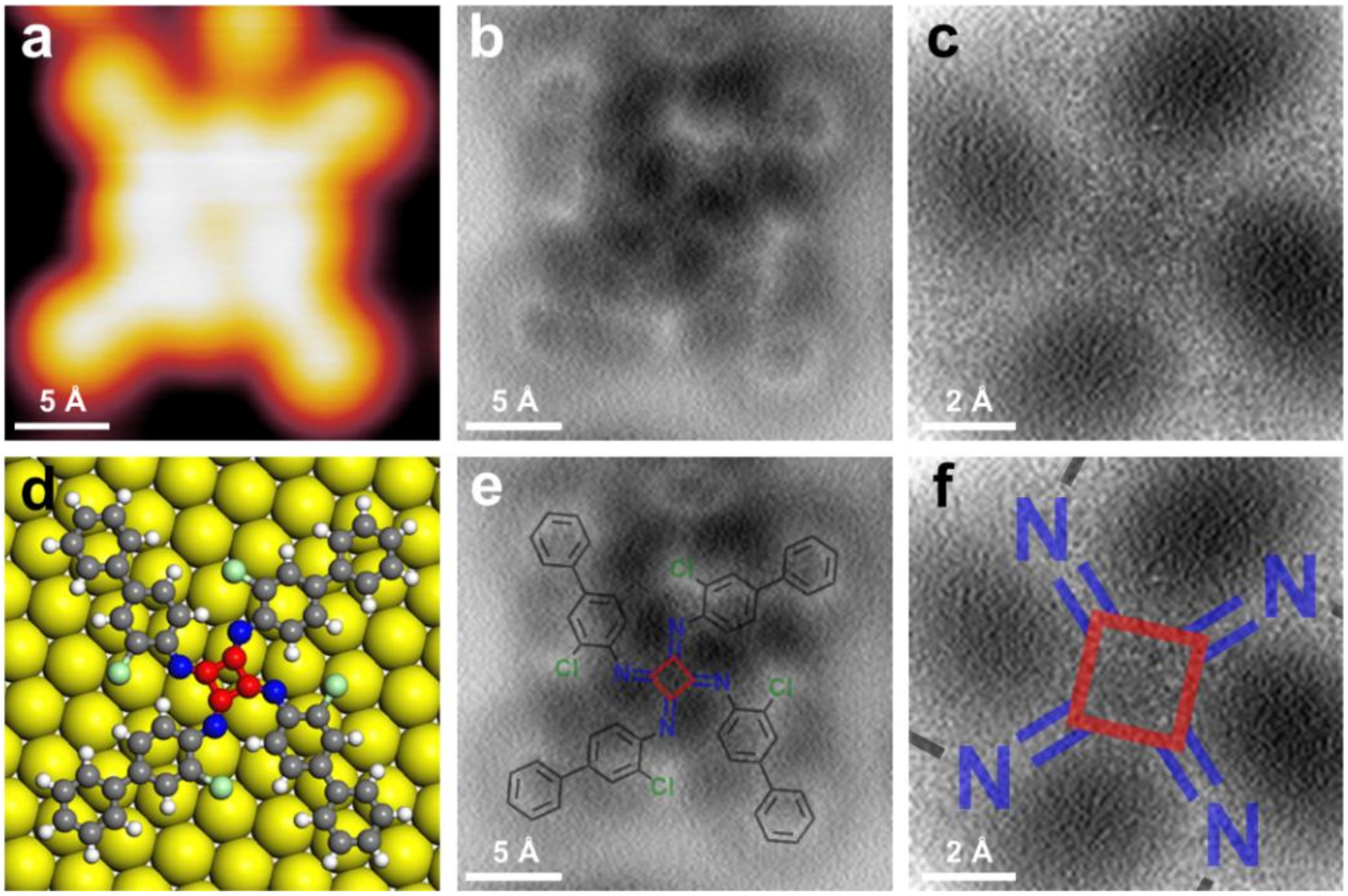

**Figure 3.** Structural characterization of tetraaza[4]radialene T1 on Au(111). (a) High-resolution STM and (b,e) nc-AFM images of an individual *R*-T1, (c,f) the corresponding zoom-in nc-AFM image of a four-membered carbon ring, and (d) structural model. Scanning parameters: (a) $U$ = 0.1 V, $I$ = 10 pA. (b, c, e, and f) Tip height: z = 226 pm with respect to STM setpoint of 100 mV, 10 pA on Au(111).

To further determine the chemical structure of T1 formed on the Au(111) surface, nc-AFM measurements were performed (Figures 3 and S4). High-resolution STM image (Figure 3a) of an individual T1 exhibits the four-leafed structure, and the corresponding nc-AFM image (Figure 3b,e) demonstrates that each leaf is Cl-substituted biphenyl derived from the precursor Cl-ICBP. Moreover, bond-resolution nc-AFM images (Figure S4a-c) at different tip heights are consistent with the corresponding nc-AFM simulations (Figure S4d-f) based on DFT-optimized model (Figure 3d), determining the chemical structure of T1. Most importantly, zoom-in nc-AFM image (Figure 3c,f) characterizes the four-membered carbon ring structure of T1, exhibiting a planar geometry. In addition, the molecular weight of tetraaza[4]radialene T1 was confirmed using time-of-flight secondary ion mass spectrometry (ToF-SIMS) — experimental results are in good agreement with theoretical predictions (Figure S5).[45,61]

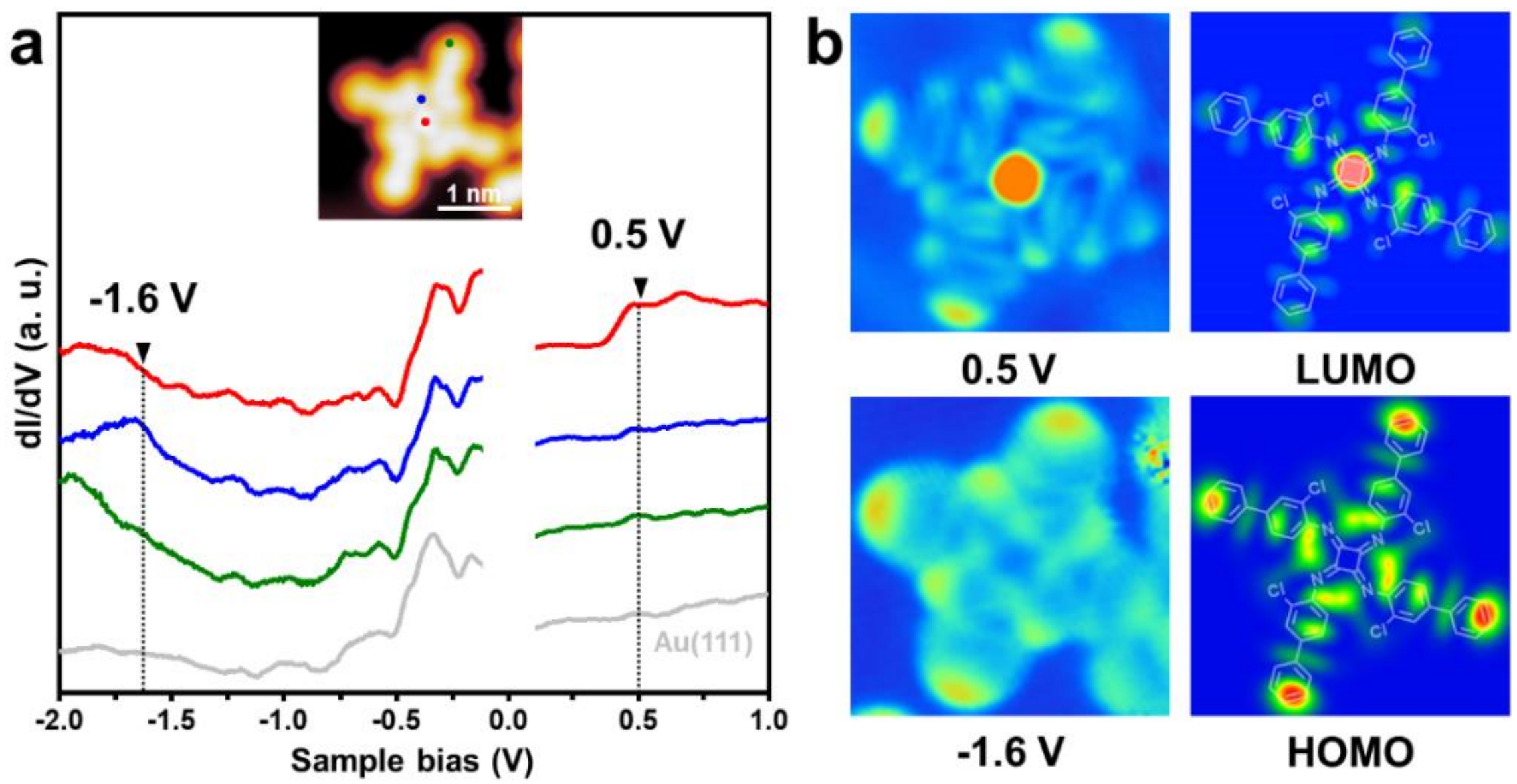

**Figure 4.** Electronic structure of tetraaza[4]radialene T1 on Au(111). (a) dI/dV spectra recorded on different locations (marked in the inset STM image, $U = 1.0$ V, $I = 20$ pA) of an individual *R*-T1. (b) Left column: constant-current dI/dV maps conducted at the energies of -1.6 and 0.5 V states. Right column: calculated LDOS of LUMO and HOMO of a free-standing tetraaza[4]radialene.

### Electronic structure of tetraaza[4]radialene on Au(111)

To characterize the electronic structure of tetraaza[4]radialene, differential conductance spectroscopy (dI/dV) was performed. As shown in Figure 4a, two resonance states were identified at -1.6 and 0.5 V on the backbone region. To avoid the topographic effects, dI/dV mapping was conducted in constant-current mode, enabling a rational comparison with the calculated local density of states (LDOS). The state at -1.6 V was found to be distributed over the biphenyl skeleton, whereas the state at 0.5 V was located at the four-membered ring (Figure 4b, left column). These features resemble the LDOS maps (Figure 4b, right column and S6) of the highest occupied molecular orbital (HOMO) and lowest unoccupied molecular orbital (LUMO). Therefore, the two states at -1.6 and 0.5 V can be assigned as the HOMO and LUMO of tetraaza[4]radialene.

### Reaction pathways from isocyanides to tetraaza[4]radialenes

To clarify the reaction mechanism from isocyanides to tetraaza[4]radialenes with homotactic configurations on Au(111), we conducted DFT calculations for the transformation process involving the stepwise and synergistic formation of four C–C bonds (Figure 5 and S7-9), in which a simplified model molecule 1-chloro-2-isocyanobenzene (Cl-ICB) was used. The initial states (IS) of proposed reaction pathways have the same structure: four Cl-ICB molecules self-assembled in a homotactic configuration. DFT calculations predict that the *o*-chlorobenzene moiety of each Cl-ICB molecule in the IS has a close-to-planar adsorption structure on Au(111), and each isocyano group is tilted with the terminal carbon atom closest to the Au atoms of the Au(111) surface (Figure 5, IS). Notably, the flat conformation and homotactic configuration can facilitate the C–C coupling reactions between isocyano group of different isocyanides.

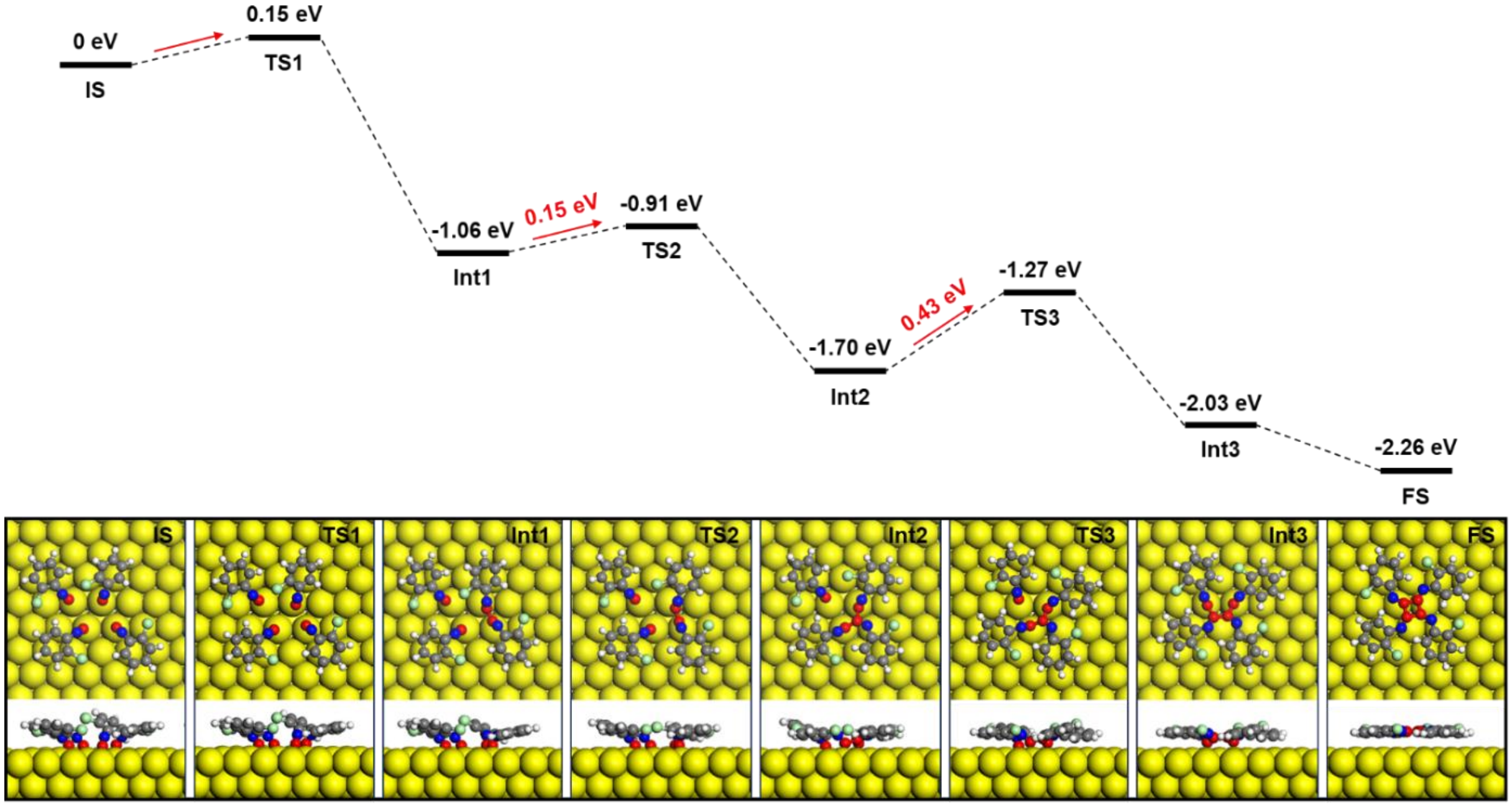

**Figure 5.** Stepwise reaction pathway from isocyanides to [4]radialenes on Au(111) involving the one-by-one formation of four C–C bonds. The corresponding calculated molecular structure, including top and side views of the initial (IS), transition (TS), and intermediate (Int) states of the reactions are shown below the energy diagrams. The red, blue, and green ball indicate the carbon and nitrogen of isocyano and Cl atom.

In the stepwise pathway for forming tetraaza[4]radialenes from isocyanides (Figure 5), two adjacent Cl-ICB molecules first undergo C–C coupling on Au(111) to form a surface-bound cumulenic intermediate (Int1), where the calculated barrier is 0.15 eV. Subsequently, Int1 reacts with one of the adjacent Cl-ICB molecules to form a surface-bound trimeric intermediate (Int2) with a similar reaction barrier. Int2 reacts with the last adjacent Cl-ICB molecule to form surface-bound ring-opening tetrameric intermediate (Int3) with a calculated reaction barrier of 0.43 eV. Finally, Int3 undergoes spontaneous intramolecular ring-closing to generate the *R*-tetraaza[4]radialene (FS). It should be noted that the stepwise [1+1+1+1] cycloaddition is an exothermic process with an overall energy decrease of approximately 2.26 eV. The rate-determining step is the coupling reaction of Int2 with the adjacent Cl-ICB molecule to form Int3, which is consistent with the experimental results — Y-shaped trimer intermediates were frequently observed, but open-ring tetramer intermediates were scarcely observed.

In addition, the other possible reaction pathways were calculated and presented in Figures S7-S9. In the other stepwise reaction pathways shown in Figures S7 and S8, the highest energy barrier for the formation of C–C bond is 0.47 eV. In contrast, the DFT-calculated synergistic reaction pathway (Figure S9) shows that the barrier for the one-step formation of four C–C bonds is 0.27 eV. Therefore, the calculated barriers for the formation of C–C bonds are low, demonstrating that both stepwise and synergistic pathways are plausible under the growth conditions.

**Formation of 2D homochiral tetraaza[4]radialene crystals via enantioselective molecular recognition**

To control the chiral assembly of tetraaza[4]radialenes to form 2D homochiral crystals on the Au(111) surface, we further design and synthesize the precursor 3,4'-dichloro-4-isocyano-1,1'-biphenyl ($Cl_2$-ICBP, Figure S1), which possesses two Cl groups at two different benzene rings. This precursor is expected to undergo the [1+1+1+1] cycloaddition on Au(111) to form tetraaza[4]radialenes (T2) with homotactic configurations because of the steric hindrance of Cl groups at the *ortho*-position of isocyano group. Moreover, the DFT-calculated electrostatic potential map of *S*-tetraaza[4]radialene (Figure 1b, inset) shows that the lowest electron density is located at the C–Cl bond axis and nitrogen atoms, and the highest electron density is located at the H atoms of C–H bonds, which may dominate the 2D molecular crystal formation on the confinement surfaces. As a result, *S*-tetraaza[4]radialenes may assemble into 2D homochiral tetraaza[4]radialene crystals on Au(111) through enantioselective molecular recognition driven by multiple C–H···Cl hydrogen-bonding interactions (Figure 1b).

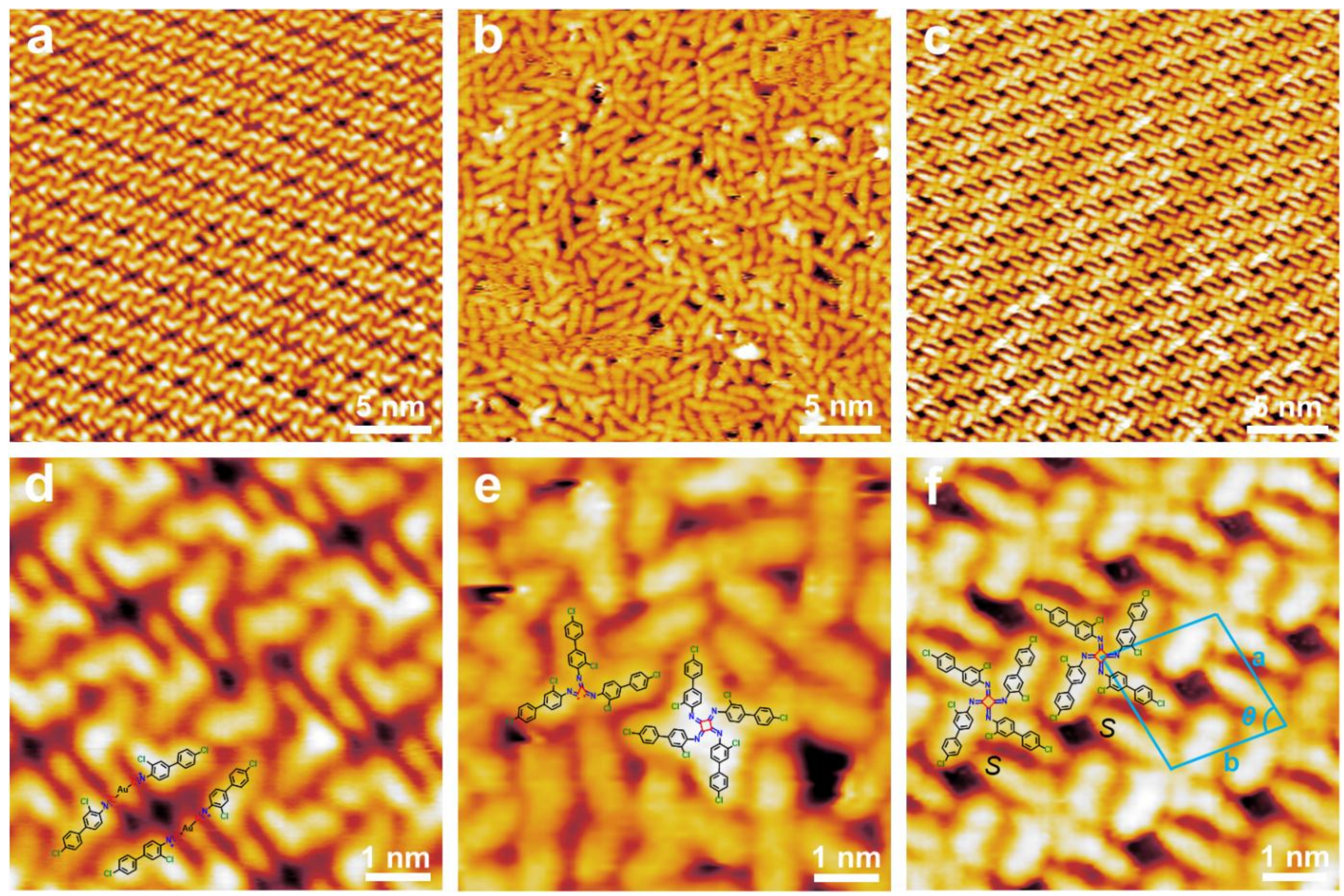

**Figure 6.** Formation of 2D homochiral tetraaza[4]radialene T2 crystals via a combination of [1+1+1+1] cycloaddition of isocyanides and enantioselective molecular recognition driven by C–H···Cl hydrogen-bonding interactions. (a and d) High-resolution and zoom-in STM images of coordination assembly of $Cl_2$-ICBP on Au(111) at room temperature. (b and e) High-resolution and zoom-in STM images after annealing at 393 K (c and f) High-resolution and zoom-in STM images of 2D homochiral (*S*)-T2 crystal formed after annealing at 433 K. Scanning parameters: (a) $U$ = 0.47 V, $I$ = 0.09 nA. (b and e) $U$ = -1.74 V, $I$ = 0.12 nA. (c and f) $U$ = -0.81 V, $I$ = 0.12 nA. (d) $U$ = 70 mV, $I$ = 0.09 nA.

After depositing the $Cl_2$-ICBP precursor onto the Au(111) surface held at approximately 293 K, STM measurements reveal the formation of large-scale regular phases (Figure 6a). High-resolution STM image (Figure 6d) shows that these regular phases consist of two $Cl_2$-ICBP molecules and one central Au adatom. Subsequently, the sample was annealed at 393 K, and most of the regular structures disappear (Figure 6b), generating an irregular phase composed of some new oligomers, such as Y-shaped trimers and various tetramers. Zoom-in STM image (Figure 6e) shows a regular tetramer, which can be assigned to the tetraaza[4]radialene (T2) with homotactic configuration.

With prolonged annealing time and the increasing of annealing temperature up to 433 K, large-scale ordered assembly domains appear (Figures 6c and S10), which consist of the homochiral tetraaza[4]radialenes (*S*-T2 or *R*-T2). Zoom-in STM images show that the unit cell parameters are $a = b = 1.83 \pm 0.05$ nm and $\theta = 87 \pm 5°$ (Figures 6f and S10c), which matches well with the corresponding STM simulations based on DFT-calculated model (Figure S10d). These results indicate that multiple intermolecular C–H···Cl hydrogen bonds can improve the intermolecular interactions, thereby guiding the enantioselective molecular recognition of tetraaza[4]radialenes.

## Conclusions

In summary, we reported the first [1+1+1+1] cycloaddition of isocyanides on the Au(111) surface to synthesize the tetraaza[4]radialenes with high chemo- and stereoselectivities. We further achieved the 2D homochiral crystallization of tetraaza[4]radialenes via multiple Cl group-directed enantioselective molecular recognition. By the combination of STM/STS and nc-AFM, we determined the atomic structure and molecular orbitals of tetraaza[4]radialene, confirming that its four-membered rings adopt a planar geometry with a localized LUMO. DFT calculations suggest that the [1+1+1+1] cycloaddition proceeds involving stepwise formation of C–C bonds and the high selectivity ascribes to the spatial steric hindrance on the Au(111) surface. Our findings provide new insights into the selective formation of conjugated rings on surfaces and have implications for engineering 2D homochiral molecular crystallization.

## Methods

### Synthesis of precursors

Precursor Cl-ICBP was synthesized from *N*-(4-bromo-2-chlorophenyl)formamide by previous reported method[59] and $Cl_2$-ICBP was synthesized from *N*-(4-bromo-2-chlorophenyl)formamide by solution methods (the detailed synthetic procedure in Supporting Information and Figure S1).

### Sample preparation and STM/nc-AFM measurement

Single crystalline Au(111) surface was cleaned by cycles of $Ar^+$ sputtering and annealing under a UHV condition (base pressure $2 \times 10^{-10}$ mbar). Precursor Cl-ICBP was evaporated from a quartz crucible onto the Au(111) surface held at the indicated temperature, and the sublimation temperature was approximately 20 °C. Then, the progressive annealing at the evaluated temperatures was performed. STM measurements were performed on a variable-temperature scanning tunneling microscope (STM; SPECS, Aarhus 150) at approximately 120 K and a Scienta Omicron Polar-STM/AFM combined system operated at approximately 4.5 K with an electrochemically etched tungsten tip. The STM images were taken in the constant-current mode using a tungsten tip and the voltages refer to the bias on the samples concerning the tip. nc-AFM images were acquired with a low-temperature scanning tunneling microscope (Scienta Omicron POLAR-STM/AFM combined system) operated at approximately 4.5 K. The qplus tip was modified with a single CO molecule at the tip apex. The oscillation amplitudes of all nc-AFM images are 30 pm.

**TOF-SIMS measurement**

The surface spectrometry was performed on a TOF.SIMS 5-100 instrument (ION-TOF GmbH, Münster, Germany) using a 30 kV $Bi_3^{++}$ as analysis beam with an incident angle of 45°. The beam current was 0.7 pA, and the field of view was $500 \times 500$ $\mu m^2$ with raster size of $128 \times 128$ pixels.

**DFT calculations**

The calculations were carried out in the framework of DFT by using the Vienna ab initio Simulation Package (VASP).[62,63] The projector augmented-wave method was used to describe the interaction between ions and electrons.[64,65] We used the generalized gradient approximation with Perdew–Burke–Ernzerhof formulism to treat exchange–correlation interaction,[66] and van der Waals (vdW) interactions were considered by using the DFT-D3 developed by Grimme.[67] The energy cutoff for the plane wave basis sets is 400 eV, and the energy between two consecutive self-consistent steps was set as $10^{-5}$ eV, respectively. The Au(111) surface was modelled by two-layered slabs separated by at least 15 Å of vacuum. For all adsorption configurations, we first performed calculations with a Γ-point only k-point sampling, and for the most stable adsorption configurations, calculations with a $1 \times 3$ k-point sampling were performed to ensure full convergence of adsorption energies.

The simulations of the reaction barriers in the manuscript are performed with the climbing image nudged elastic band (CI-NEB) method for finding saddle points and minimum energy paths (http://theory.cm.utexas.edu/vtsttools/neb.html), which is at T = 0 without including entropy and vibrations.[68] The STM simulation was performed with the Tersoff-Hamann method.[69–71]

**Acknowledgements**

This work was supported by the National Natural Science Foundation of China (Nos. 22272050 and 92580139), the Shanghai Municipal Science and Technology Qi Ming Xing Project (No. 22QA1403000), the Shanghai Postdoctoral Excellence program (2025646), and the Fundamental Research Funds for the Central Universities. We are grateful for the TOF-SIMS measurement support from Vacuum Interconnected Nanotech Workstation (Nano-X).

**Author contributions**

D.-Y.L. conceived the experiments; J.-W.L. and W.-W.G. performed on-surface synthesis experiments with the supervision of P.-N.L. and D.-Y. L.; Y.W. and J.-H.F. performed nc-AFM and STS characterizations with the supervision of P.-N.L. and D.-Y. L.; J.-W.L. and L.-X.K. conducted theoretical computations with the supervision of P.-N.L. and D.-Y. L.; J.-W.L, Y.W., and D.-Y.L. analyzed the data; J.-W. L., Y.W., and D.-Y.L. wrote the manuscript; All authors discussed the results and helped write the manuscript at all stages.

## Competing interests

The authors declare no competing interests.

## Additional information

Correspondence and requests for materials should be addressed to Ying Wang, Pei-Nian Liu, and Deng-Yuan Li.